# Computational Studies about the Interactions of Nanomaterials with Proteins and their Impacts[*]

An De-Yi(安德义)[a)b)], Su Ji-Guo(苏计国)[b)], Li Chun-Hua(李春华)[c)†], and Li Jing-Yuan(李敬源)[a)‡]

[a)]*CAS Key Lab for Biomedical Effects of Nanomaterials and Nanosafety, Institute of High Energy Physics, Chinese Academy of Science, Beijing* 100049, *China*

[b)]*College of Science, Yanshan University, Qinhuangdao* 066004, *China*

[c)]*College of Life Science and Bioengineering, Beijing University of Technology, Beijing* 100124, *China*

Intensive concerns about the biosafety of nanomaterials demand the systematic study of the mechanisms about their biological effects. Many biological effects can be attributed to the interaction of nanomaterials with protein and their impacts on protein function. On the other hand, nanomaterials exhibit the potential in a variety of biomedical applications, many of which also involve the direct interaction with protein. In this paper, we review some recent computational studies about this subject, especially the interaction of carbon and gold nanomaterials. Besides the hydrophobic and π-stacking interactions, the interaction mode of carbon nanomaterials can be regulated by their functional groups. And the coating of gold nanomaterials also adjusts their interaction mode, in addition to the coordination interaction with cysteine's sulfur group and histidine's imidazole group. Moreover, nanomaterials can interact with multiple proteins and the impacts on protein activity are attributed to a wide spectrum of mechanisms. The findings about the mechanisms of nanomaterial-protein interaction can further guide the design and development of nanomaterial to realize the applications in disease diagnosis and treatment.

[*] Project supported by the National Natural Science Foundation of China (Grant Nos. 21273240, 11204267 and 11474013).

[‡]Corresponding author. E-mail: lijingyuan@ihep.ac.cn
[†]Corresponding author. E-mail: chunhuali@bjut.edu.cn

## 1. Introduction

Because of their small size and/or nanostructure, nanomaterials often have large specific surface area and produce quantum effects.[1,2] Both endow nanomaterials with a variety of unique physical and chemical properties,[3-5] leading to the application in different areas including healthcare, electronics, cosmetics and textiles. In the meantime, intensive attentions have been paid to the biological safety of nanomaterials.[6,7] However, the potential threats of nanomaterials to human health especially the corresponding mechanisms largely remain elusive. On the other hand, nanomaterials exhibit promising potential in biomedical researches, e.g. drug delivery, diagnosis, bioimaging, and therapeutic agents.[8-14] Many biological effects of nanomaterials (both positive and negative) are related to their impacts on the structure and function of protein. Hence, the studies about the interaction with protein and subsequent impact on protein function are crucial to a better understanding of the biological effect of nanomaterial. However, the metabolic processes of nanomaterials are very complicated in nature. For example, various proteins can adsorb on the surface of nanomaterial and form protein corona. The composition and content of protein within corona are very complicated and keep developing.[15] Hence, it is very difficult to elucidate the underlying mechanism of a given biological effect. On top of this, relevant studies were often hampered by the limitation of spatial and temporal resolution of experimental techniques.[16]

With the development of force field and the improvement of sampling efficiency as well as computing power, computational methods are widely applied to study a variety of biological processes and drug designs. They have also been proven to be powerful in the research about the biological effect of nanomaterials. Based on computational studies, the interaction between nanomaterials and proteins and the impacts on the structure and function of biological molecules can be effectively identified. Moreover, these findings of computational studies can even guide the design and development of nanomaterial in the biomedical applications including disease diagnosis and treatment.

## 2. Carbon nanomaterial

Carbon nanomaterial is one of the most important inorganic nanomaterials, which includes graphene, fullerene, carbon nanotube, etc. A variety of carbon nanomaterials exhibit diversified bioapplications, such as drug and gene delivery, contrast agents, therapeutic agents, and components of biosensors.[17-21] On the other hand, the inhaled carbon nanomaterials may lead to stress, inflammation, lung insult and a variety of cardiovascular effects.[22-28] There are growing numbers of computational studies to investigate the mechanisms behind the spectrum of biological effects. Because of their well-defined structure, carbon nanomaterials also serve as the representative hydrophobic nanomaterials in the pioneering studies about the protein-nanomaterial interaction. To our knowledge, one of the first simulation works on the protein-nanomaterial interaction is to study the binding of fullerene C60 to antibody.[29] Jianpeng Ma and co-workers found C60 interacts with antibody and forms the complex with high affinity and specificity. The binding is attributed to shape complementarity and extensive side-chain interaction, including hydrophobic and π-stacking interactions. Since such binding mode is similar to many other protein–ligand complexes, C60 may competitively interact with protein binding site and disturb the protein-ligand recognition.

## 2.1 Fullerene derivative, Gd@C$_{82}$(OH)$_{22}$

Fullerene is the carbon cage with the diameter of around 1 nm. It has been widely exploited in the fields of bioimaging, drug delivery and antitumor therapy,[30-32] while the poor aqueous solubility of fullerene poses challenges to their further applications. And there have been a variety of fullerene derivatives, for example, fullerene can be hydroxylated to fullerenol. Surface modification not only improves their solubility, but also endows additional feature to fullerenes. Moreover, metal ion can be embedded in the carbon cage of fullerenol to form metallofullerenol. In general, metallofullerenol shares similar surface properties with fullerenol, while the embedded metal atom gives rise to unique properties of metallofullerenol,[33-36] as will be discussed below.

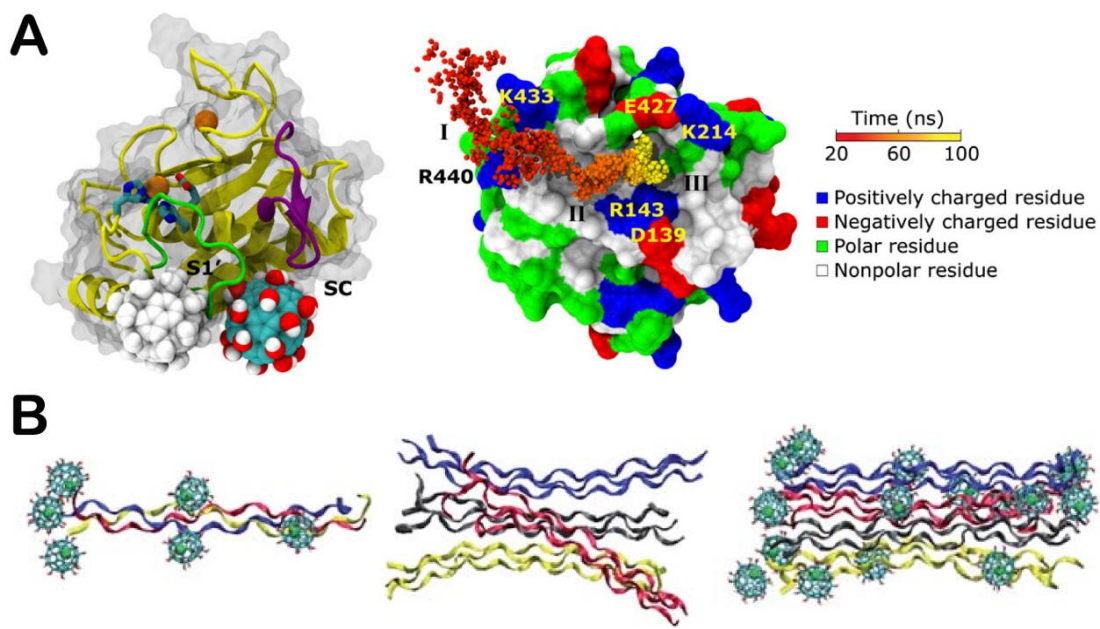

Figure 1. Interactions of Gd@C$_{82}$(OH)$_{22}$ with proteins. (A, left) Representative binding mode (solid ball) as well as alternative mode (gray ball) of Gd@C$_{82}$(OH)$_{22}$. (right) the binding dynamics can be characterized as three phases. Adapted with permission from Ref. [40]. Copyright 2012 National Academy of Sciences, USA. (B, left) Representative snapshot of the tropocollagen molecule bound by Gd@C$_{82}$(OH)$_{22}$. Snapshot of the tropocollagen tetramer (middle) and the tetramer bound by Gd@C$_{82}$(OH)$_{22}$ (right). Reproduced from Ref. [46] with permission from The Royal Society of Chemistry.

Experimental studies have shown that a typical endohedral metallofullerenol, Gd@C$_{82}$(OH)$_{22}$ can effectively inhibit tumor growth with low toxicity both *in vivo* and *in vitro*.[37,38] After the treatment of Gd@C$_{82}$(OH)$_{22}$ the thickness and flexibility of the fibrous layer surrounding tumor significantly increase, leading to the formation of a fibrous cage to imprison the tumor tissue and prevent its metastasis. The impact on the fibrous layer thickness can be largely attributed to the suppression of the expression of matrix metalloproteinases (MMPs) as well as the reduction of their activities.[39,40] In order to elucidate this important mechanism involved in the antitumor effect of metallofullerenol, Ruhong Zhou and co-workers studied the interaction of Gd@$C_{82}$(OH)$_{22}$ with MMP-9 and its impact on the protein function.[40] Metallofullerenol can firmly bind to protein, while does not disturb its structure. Interestingly, instead of direct interaction with catalytic $Zn^{2+}$ ion, Gd@$C_{82}$(OH)$_{22}$ was found to allosterically modulate the S1' ligand-specificity loop which might interfere with the incoming substrate. In other words, metallofullerenol inhibits the activity of MMP-9 in an indirect way. Moreover, the authors successfully characterized the complete binding process as three stages, and identified the critical interaction of each stage.

In addition to its impacts on the fibrous layer thickness, Gd@$C_{82}$(OH)$_{22}$ was found to affect the structure and biophysical properties of collagen fiber,[39] the major component of fibrous layer.[41-45] The fibrous layer become softer after the treatment, and this is also an important antitumor mechanism of metallofullerenol. Given by this, we investigated the interaction of Gd@$C_{82}$(OH)$_{22}$ with molecular collagen (tropocollagen molecule), and its impact on the structure and assembly of protein.[46] Gd@$C_{82}$(OH)$_{22}$ can strongly bind to tropocollagen, largely due to the hydrogen bond interaction with protein. Adhered Gd@$C_{82}$(OH)$_{22}$ can enhance the stability of native triple helical structure of tropocollagen and facilitate the protein assembly. Interestingly, the interaction of nanoparticle with protein was often considered to disturb protein structure or induce abnormal assembly,[47-50] while, as indicated in our work, nanoparticle may also enhance the native structure and assembly of protein. In the early stage of collagen fiber formation, Gd@$C_{82}$(OH)$_{22}$ can form hydrogen bonds with multiple tropocollagen molecules acting as a "fullerenol-mediated bridge" and enhance the interaction among collagen molecules over the course of fiber nucleation.[51-56]. While, during the growth of collagen fiber metallofullerenol may interfere the interactions among proteins and affect the structure and stiffness of collagen fibril layer.

## 2.2 Carbon nanotube, CNT

Carbon nanotube (CNT) is another widely used carbon nanomaterial,[57] it shares similar challenges (e.g. poor solubility) and potential as the fullerene counterparts, such as drug design, drug delivery, tumor therapy, tissue engineering, DNA recognition, and biosensor design.[58-62] Besides, CNT is often used as a representative hydrophobic nanoparticle to study the role hydrophobic interaction plays in the protein-nanoparticle interaction, and the impact on the protein structure and function.

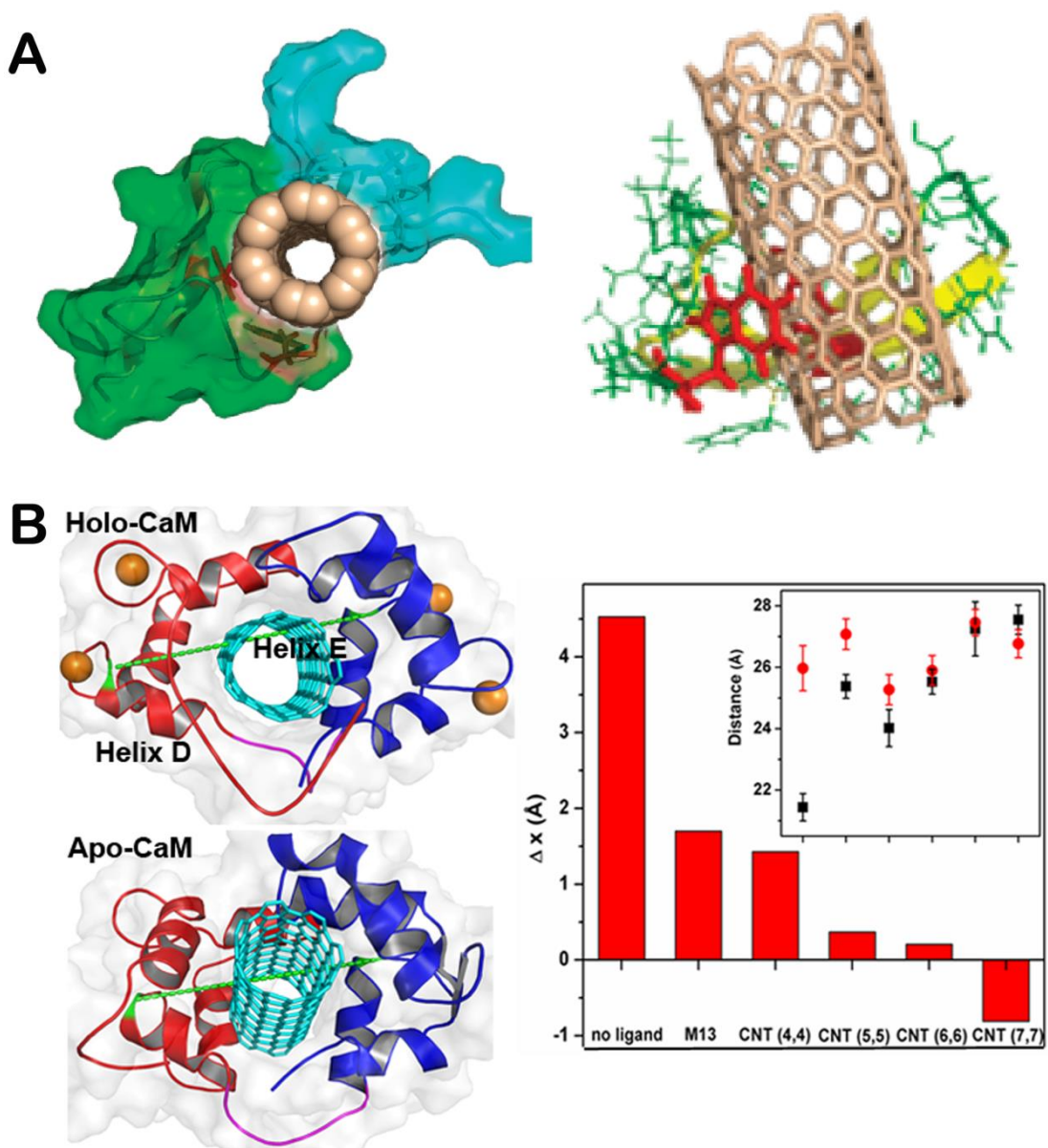

Figure 2. Impacts of CNTs on protein function. (A, left) The binding of SWCNT hinders the interaction between YAP65 (green) and its native ligand PRM (navy). (right) The interaction mode of SWCNT. The binding scaffold residues are highlighted as red sticks. Adapted with permission from Ref. [48]. Copyright 2010 American Chemical Society. (B, left) Representative configuration of CNT-CaM complex and the definition of the end-to-end distance of inter-domain linker (right) Difference in the end-to-end distance between two states. Inset: End-to-end distances of inter-domain linker in different complexes. Reproduced from Ref. [68] with permission from The Royal Society of Chemistry.

Ruhong Zhou and co-workers took WW domains (i.e. YAP65, YJQ8, and PIN1) as examples to study the interaction of CNT with protein and the subsequent impact on the protein activity.[48] As signaling and regulatory proteins, WW domains can identify and bind to the proline-rich motifs (PRMs).[63-67] The authors found CNT can plug into hydrophobic core of WW domains because of the interaction with hydrophobic residues. More importantly, the binding of CNT blocks PRM active site and thus hinders the interaction of PRM with WW domain.

Besides direct blockage of protein active site, the impact of CNT on protein function can be attributed to other mechanisms. Recently,[68] we investigated the binding of CNT to calmodulin (CaM) and its impact on $Ca^{2+}$-dependent dynamic properties of CaM. In addition, the size dependence of CNT's biological effect was studied. CaM plays a crucial role in the calcium signal transduction pathway.[69] It can bind to a large variety of enzymes in a $Ca^{2+}$-dependent manner:[70,71] $Ca^{2+}$ facilitates ligand binding by enhancing hydrophobic interactions between ligand and protein;[72-75] and the ion removal triggers ligand dissociation. We found CNTs can recognize the hydrophobic binding pocket of CaM. While small CNT shows a similar behavior to the native substrate M13 peptide in its dissociation from $Ca^{2+}$-free CaM, wider CNTs keep binding to CaM in the absence of $Ca^{2+}$, indicating a potential failure of $Ca^{2+}$ regulation and the inhibition of calcium-dependent signal transduction pathway. Such size-dependent impact on protein dynamic properties is largely due to the fact that the hydrophobic interaction of wider CNTs with protein is strong enough and does not require the assistance of $Ca^{2+}$. Hence, the binding of wider CNT can dodge the regulation of $Ca^{2+}$. The simulation results about the binding of CNT and the failure of $Ca^{2+}$ regulation were further confirmed by circular dichroism spectroscopy.

2.3 Graphene
Graphene is a 2D plate-like carbon material with a series of extraordinary structural, mechanical and electronic properties.[76] There are growing numbers of researches to explore the biomedical applications of graphene and its interaction with biological system.[77] Zuo *et al.* used molecular dynamics simulation to investigate the adsorption of protein villin headpiece (HP35) onto graphene.[78] HP35 protein is composed of three-helix bundle and most of its native secondary and tertiary structures change after the adsorption. The adsorption stability is largely attributed to the π-stacking interaction between graphene and the aromatic residues of protein. Moreover, because of its softness the shape of graphene can adapt to the distribution of aromatic residues and form strong π-stacking with protein. The authors also compared with the interaction between HP35 and CNT as well as C60, and found that the surface curvature of nanomaterials with same chemical component can affect their interaction mode with proteins. In addition to the study about protein adsorption, the interaction between graphene and phospholipid, another kind of important biological molecule, has attained intensive attention. Zhou and co-workers found the graphene can extract phospholipids from membrane and destroy membrane structure.[79] Meanwhile, Gao and co-workers systematically investigated the insertion and entry of graphene into membrane.[80]

**3. Gold nanomaterial**
Noble metal nanomaterials possess a variety of unique physical properties, they can serve as typical systems to study the quantum confinement effect. Because of their stable chemical property and high biocompatibility, gold nanomaterials become one of the most popular noble metal nanomaterials. With the rapid progress in the synthesis and modification of gold nanomaterial, there have been a wide spectrum of gold nanomaterials including nanorods, nanoclusters, nanobelts, nanostars and polyhedral nanoparticles. The distinct optical properties of gold nanomaterials engender their potential applications in the biomedical imaging, sensing, as well as photothermal therapy. And the studies about the interaction of gold nanomaterials with biological molecules are highly demanded.

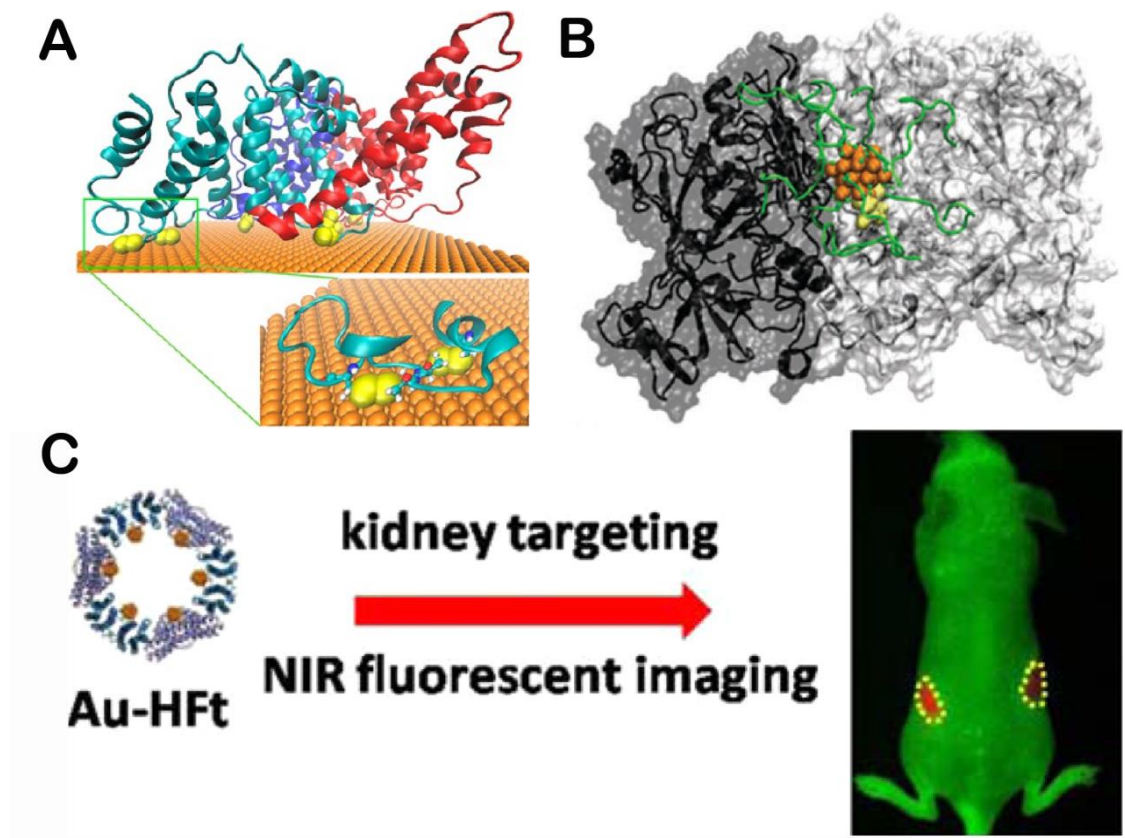

Figure 3. Gold nanoparticles interacting with proteins. (A) The binding of BSA to Au (111) surface of AuNRs. Three domains of BSA are colored in cyan, red, blue; and disulfides are highlighted as yellow spheres. Adapted with permission from Ref. [92]. Copyright 2013 American Chemical Society. (B) The AuNC–TrxR1 complex. The gold atoms, peptide chain, Cys/Sec residues of active site are in orange, green and yellow respectively. Reproduced from Ref. [112] with permission from The Royal Society of Chemistry. (C) Designed near-infrared (NIR) fluorescent hybrid nanocomposite: multiple AuNCs within the cage of H-ferritin (HFt). This designed probe can realize kidney targeting and NIR imaging of live animal. Adapted with permission from Ref. [110]. Copyright 2015 American Chemical Society.

**3.1 Gold nanorod**
The sizes of gold nanorod, AuNR, are 10-20 nm in width and 10-100 nm in length. The optical adsorption of AuNR can be effectively regulated by changing its aspect ratio.[81] The optical properties of gold nanorod form the basis for biomedical applications, and their potential impacts on living systems are then attracting much attention.[82,83] The protein corona assembled on the surface of AuNR during its metabolic process modulates the biological responses by mitigating the cytotoxicity,[84-87] changing the biodistribution[88-90] and flammatory response of AuNR.[91] As the most abundant serum protein, Bovine Serum Albumin (BSA) is the important component of corona. The study about its adsorption on AuNR is critical to understand the formation of protein corona. Previous experimental studies have shown that BSA can stably bind to the surface of AuNR and improve its biocompatibility,[92] while the detailed molecular mechanism and the corresponding structure still remain elusive. We combined the experimental synchrotron radiation (SR)-based analytical techniques[93-97] with molecular dynamics simulation to investigate the interaction mode of BSA corona with AuNR. It has been well accepted that the binding of BSA is largely attributed to the Au-S coordination between the gold atoms and sulfur atoms of cysteine residues. There are 17 disulfide bonds between cysteine residues in BSA, most of which are exposed to solvent. We found there are 8 disulfide bonds distributed around a plane, denoted to *plane S*. The adsorption process of BSA was further investigated by molecular dynamics simulation. The *plane S* can serve as the binding interface, and there are at least 12 Au-S coordination bonds form during the adsorption. These findings were confirmed by SR S K-edge X-ray absorption near-edge structures (XANES)[98] and SR-based microbeam X-ray fluorescence[99-101] (XRF) results. In addition, our study proposes such combined approach can effectively study the interfacial interactions of protein corona with AuNR, which should improve the understanding about the nature of corona's protective effects.

**3.2 Gold nanocluster**
Gold nanocluster, AuNC, is composed of a few to roughly a hundred atoms with the diameter below 2 nm. And the properties of AuNC are distinct to both isolated atoms and larger nanoparticles.[102] As a representative fluorescent nanomaterial, AuNC becomes attractive to biolabeling and bioimaging applications owing to its ultra-small size, nontoxicity and highly fluorescent properties.[103-105] In order to improve their suspension stability, AuNCs are often protected with various coatings such as alkanethiols, DNAs, peptides or even within protein cage,[106-109] resulting in the bioinorganic hybrid nanomaterials. The coatings can also modulate the surface properties of AuNC and improve its biocompatibility.

As indicated in the previous work of our collaborators, ferritin protein cage can guide the formation of gold nanoclusters, leading to the gold cluster-based hybrid nanocomposite.[110] Ferritin complex is composed of 24 monomers with two types, i.e. heavy chain and light chain (H- and L-ferritin, respectively). According to our simulation result, the nucleation sites of gold clusters locate in His-rich surface region of H-ferritin, HFt. The interaction of gold atoms with imidazole group of histidine facilitates the formation of gold cluster. Accordingly, our collaborators used the protein nanocage composed of all HFts to synthesize the hybrid nanocomposite containing 24 AuNCs. The number of nanoclusters within nanocage was confirmed by cryo-EM image. Moreover, such Au-HFt nanocomposite can serve as near infrared (NIR) probe with high fluorescent yield, showing powerful tissue penetrating ability.

Because of the kidney targeting ability of ferritin, this designed NIR Au-HFt probe may be exploited in kidney disease diagnosis of live animal.

More interestingly, AuNC can directly bind to target protein and affect the protein activity, exhibiting the potential in disease treatment. Our collaborator synthesized peptide coated $Au_{25}$ cluster,[111] and found the AuNC can specifically bind to Thioredoxin Reductase1, TrxR1.[112] TrxR1 is important to the regulation of cellular redox level and often overexpressed in cancer cells.[113,114] The protein has been well recognized as the potential target for the anti-tumor therapeutic agent.[115,116] The treatment of peptide-coated AuNC was found to effectively suppress the protein activity, resulting in the increased concentration of reactive oxygen species and the subsequent apoptosis of tumor cells. The potential of peptide-coated AuNC in tumor therapy requires the detailed study about the corresponding mechanism especially the interaction mode of AuNC with protein. We firstly used molecular docking method[117] to search the binding site throughout the surface of TrxR1 and successfully identified a putative binding region around the active site Cys497-Sec498 (Sec, selenocysteine). Additional molecular dynamics simulation was then performed to assess the binding stability of AuNC. In general, the AuNC can directly bind to the region around active site, mainly due to electrostatic attraction between positively charged coating peptides and negatively charged surface residues around the active site. Besides, the hydrogen bond and hydrophobic interactions are also involved in the binding of such coated AuNC, which facilitates the subsequent coordination interaction of $Au_{25}$ cluster with cysteine and selenocysteine.[118] Taken together, such AuNCs can selectively recognize TrxR1 *in vivo*. The studies about molecular mechanism of these biological effects may be inspiring to the design of therapeutic gold nanoparticles against disease involving TrxR1, e.g. cancer.

## 4. Conclusions

In this paper, we reviewed recent computational studies about interaction of protein with carbon and gold nanomaterials. The interaction of carbon nanomaterial with protein is largely attributed to the hydrophobic and π-stacking interactions. While the binding of protein to pristine gold nanomaterial mainly results from Au-S coordination as well as the interaction with imidazole group of histidine, the gold nanomaterials are often modified by various coatings and the interaction mode of coated gold nanomaterial also involves electrostatic, hydrophobic and hydrogen bond interactions.

The interaction with nanomaterial often results in the influence on protein activity. For example, CNT can interact with the active site of WW domain and hinder the interaction with native ligand. Besides, the binding of CNT can interrupt the dynamic properties of CaM in a size-dependent manner. On the other hand, the impacts of nanomaterial on the protein structure and activity provide possible resolutions to the disease treatment. Metallofullerenol $Gd@C_{82}(OH)_{22}$ was found to effectively inhibit tumor growth by increasing the thickness and flexibility of fibrous layer to "prison" cancer cells. The anti-tumor effect of $Gd@C_{82}(OH)_{22}$ involves multiple target proteins. The expression and activity of matrix metalloproteinases are down regulated by nanoparticle. In addition, $Gd@C_{82}(OH)_{22}$ can affect the structure and assembly of molecular collagen which should lead to the reduced stiffness of collagen fibrous layer.

Because of their fast growing applications in biomedical studies, the interaction of gold nanomaterial with protein has also attained wide interests. The importance of Au-S coordination in such systems has been well recognized, and the binding behavior of some proteins depends on the distribution of exposed cysteine residues and can be captured by experimental techniques like S K-edge XANES. The binding of peptide-coated AuNC to the cysteine-containing active site of TrxR1 results in the inhibition of protein activity and the apoptosis of cancer cell. On the other hand, the imidazole group of histidine is also involved in the interaction with gold nanoparticles. For example, histidine-rich surface region of HFt can guide the nucleation and growth of AuNCs.

In short, complicated nature of metabolite process of nanomaterial poses the challenge to study their interactions with protein *in vivo*. On the other hand, computational studies can effectively investigate the mechanisms about the interaction of nanomaterial with protein and the subsequent impact on protein activity, and explain the underlying mechanisms of biological effects. As described above, nanomaterials often interact with multiple target proteins and the modulation of protein activity can be attributed to a wide spectrum of mechanisms. In addition, simulation results can further guide the design of nanomaterial to integrate functionality and/or enhance the desired properties of nanomaterial.

**References**
[1]   Ball P and Garwin L 1992 *Nature* **355** 761
[2]   Halperin W P 1986 *Rev. Mod. Phys.* **58** 533
[3]   Steigerwald M L and Brus L E 1990 *Acc. Chem. Res.* **23** 183
[4]   Wang Y 1991 *Acc. Chem. Res.* **24** 133
[5]   Weller H 1993 *Adv. Mater.* **5** 88
[6]   Wang X F, Zhu M T and Li J Y 2012 *Chin. J. Chem .* **30** 1931
[7]   Zuo G H, Kang S G, Xiu P, Zhao Y L and Zhou R H 2013 *Small* **9** 1546
[8]   Subbiah R, Veerapandian M and Yun K S 2010 *Curr. Med. Chem.* **17** 4559
[9]   Kim B Y, Rutka J T and Chan W C 2010 *N. Engl. J. Med.* **363** 2434
[10] Delehanty J B, Boeneman K, Bradburne C E, Robertson K and Medintz I L 2009 *Expert Opin.*

*Drug Deliv.* **6** 1091

[11] Choi H S and Frangioni J V 2010 *Mol. Imaging* **9** 291

[12] Janib S M, Moses A S and MacKay J A 2010 *Adv. Drug Deliver Rev.* **62** 1052

[13] Oliveira J M, Salgado A J, Sousa N, Mano J F and Reis R L 2010 *Prog. Polym. Sci.* **35** 1163

[14] Ahmad M Z, Akhter S, Jain G K, Rahman M, Pathan S A, Ahmad F J and Khar R K 2010 *Expert Opin. Drug Deliv.* **7** 927

[15] Mahmoudi M, Lynch I, Ejtehadi M R, Monopoli M P, Bombelli F B and Laurent S 2011 *Chem. Rev.* **111** 5610

[16] Dror R O, Jensen M O, Borhani D W and Shaw D E 2010 *J. Gen. Physiol.* **135** 555

[17] Cai W B and Chen X Y 2007 *Small* **3** 1840

[18] Son S J, Bai X and Lee S B 2007 *Drug Discov. Today* **12** 650

[19] Son S J, Bai X and Lee S 2007 *Drug Discov. Today* **12** 657

[20] Foldvari M and Bagonluri M 2008 *Nanomedicine* **4** 173

[21] Foldvari M and Bagonluri M 2008 *Nanomedicine* **4** 183

[22] Nel A 2005 *Science* **308** 804

[23] Nel A, Xia T, Madler L and Li N 2006 *Science* **311** 622

[24] Oberdorster G, Oberdorster E and Oberdorster J 2005 *Environ. Health Perspect.* **113** 823

[25] Ma-Hock L, Treumann S, Strauss V, Brill S, Luizi F, Mertler M, Wiench K, Gamer A O, van Ravenzwaay B and Landsiedel R 2009 *Toxicol. Sci.* **112** 468

[26] Schipper M L, Nakayama-Ratchford N, Davis C R, Kam N W S, Chu P, Liu Z, Sun X M, Dai H J and Gambhir S S 2008 *Nat. Nanotechnol.* **3** 216

[27] Shvedova A A, Kisin E R, Mercer R, Murray A R, Johnson V J, Potapovich A I, Tyurina Y Y, Gorelik O, Arepalli S, Schwegler-Berry D, Hubbs A F, Antonini J, Evans D E, Ku B K, Ramsey D, Maynard A, Kagan V E, Castranova V and Baron P 2005 *Am. J. Physiol. Lung Cell Mol. Physiol.* **289** 698

[28] Li Z, Hulderman T, Salmen R, Chapman R, Leonard S S, Young S H, Shvedova A, Luster M I and Simeonova P P 2007 *Environ. Health Perspect.* **115** 377

[29] Noon W H, Kong Y and Ma J 2002 *Proc. Natl. Acad. Sci. U. S. A.* **99** 6466

[30] Da Ros T and Prato M 1999 *Chem. Commun.* 663

[31] Montellano A, Da Ros T, Bianco A and Prato M 2011 *Nanoscale* **3** 4035

[32] Anilkumar P, Lu F, Cao L, Luo P G, Liu J H, Sahu S, Tackett K N, Wang Y and Sun Y P 2011 *Curr. Med. Chem.* **18** 2045

[33] Liang X J, Meng H, Wang Y Z, He H Y, Meng J, Lu J, Wang P C, Zhao Y L, Gao X Y, Sun B Y, Chen C Y, Xing G M, Shen D W, Gottesman M M, Wu Y, Yin J J and Jia L 2010 *Proc. Natl. Acad. Sci. U. S. A.* **107** 7449

[34] Chen C Y, Xing G M, Wang J X, Zhao Y L, Li B, Tang J, Jia G, Wang T C, Sun J, Xing L, Yuan H, Gao Y X, Meng H, Chen Z, Zhao F, Chai Z F and Fang X H 2005 *Nano Lett.* **5** 2050

[35] Yang D, Zhao Y L, Guo H, Li Y N, Tewary P, Xing G M, Hou W, Oppenheim J J and Zhang N 2010 *ACS Nano* **4** 1178

[36] Liu Y, Jiao F, Qiu Y, Li W, Lao F, Zhou G Q, Sun B Y, Xing G M, Dong J Q, Zhao Y L, Chai Z F and Chen C Y 2009 *Biomaterials* **30** 3934

[37] Wang J X, Chen C Y, Li B, Yu H W, Zhao Y L, Sun J, Li Y F, Xing G M, Yuan H, Tang J, Chen Z, Meng H, Gao Y X, Ye C, Chai Z F, Zhu C F, Ma B C, Fang X H and Wan L J 2006 *Biochem. Pharmacol.* **71** 872

[38] Meng H, Xing G M, Sun B Y, Zhao F, Lei H, Li W, Song Y, Chen Z, Yuan H, Wang X X, Long J, Chen C Y, Liang X J, Zhang N, Chai Z F and Zhao Y L 2010 *ACS Nano* **4** 2773
[39] Meng H, Xing G M, Blanco E, Song Y, Zhao L N, Sun B Y, Li X D, Wang P C, Korotcov A, Li W, Liang X J, Chen C Y, Yuan H, Zhao F, Chen Z, Sun T, Chai Z F, Ferrari M and Zhao Y L 2012 *Nanomedicine* **8** 136
[40] Kang S G, Zhou G, Yang P, Liu Y, Sun B, Huynh T, Meng H, Zhao L, Xing G, Chen C, Zhao Y and Zhou R 2012 *Proc. Natl. Acad. Sci. U. S. A.* **109** 15431
[41] Kadler K E, Holmes D F, Trotter J A and Chapman J A 1996 *Biochem. J* **316** 1
[42] Myllyharju J and Kivirikko K I 2001 *Ann. Med.* **33** 7
[43] Myllyharju J and Kivirikko K I 2004 *Trends Genet.* **20** 33
[44] Viguet-Carrin S, Garnero P and Delmas P D 2006 *Osteoporos. Int.* **17** 319
[45] Grabowska M 1959 *Nature* **183** 1186
[46] Yin X, Zhao L, Kang S G, Pan J, Song Y, Zhang M, Xing G, Wang F, Li J, Zhou R and Zhao Y 2013 *Nanoscale* **5** 7341
[47] Balamurugan K, Gopalakrishnan R, Raman S S and Subramanian V 2010 *J. Phys. Chem. B* **114** 14048
[48] Zuo G H, Huang Q, Wei G H, Zhou R H and Fang H P 2010 *ACS Nano* **4** 7508
[49] Balamurugan K, Singam E R A and Subramanian V 2011 *J. Phys. Chem. C* **115** 8886
[50] Ge C C, Du J F, Zhao L N, Wang L M, Liu Y, Li D H, Yang Y L, Zhou R H, Zhao Y L, Chai Z F and Chen C Y 2011 *Proc. Natl. Acad. Sci. U. S. A.* **108** 16968
[51] Bachmann A, Kiefhaber T, Boudko S, Engel J and Bachinger H P 2005 *Proc. Natl. Acad. Sci. U. S. A.* **102** 13897
[52] Engel J and Bachinger H P 2005 *Collagen* **247** 7
[53] Dai N, Wang X J and Etzkorn F A 2008 *J. Am. Chem. Soc.* **130** 5396
[54] Veld P J and Stevens M J 2008 *Biophys. J.* **95** 33
[55] Gurry T, Nerenberg P S and Stultz C M 2010 *Biophys. J.* **98** 2634
[56] Raman S S, Gopalakrishnan R, Wade R C and Subramanian V 2011 *J. Phys. Chem. B* **115** 2593
[57] Iijima S 1991 *Nature* **354** 56
[58] Zanello L P, Zhao B, Hu H and Haddon R C 2006 *Nano Lett.* **6** 562
[59] Prato M, Kostarelos K and Bianco A 2008 *Acc. Chem. Res.* **41** 60
[60] Bhirde A A, Patel V, Gavard J, Zhang G F, Sousa A A, Masedunskas A, Leapman R D, Weigert R, Gutkind J S and Rusling J F 2009 *ACS Nano* **3** 307
[61] Tu X M, Manohar S, Jagota A and Zheng M 2009 *Nature* **460** 250
[62] Thakare V S, Das M, Jain A K, Patil S and Jain S 2010 *Nanomedicine* **5** 1277
[63] Chen H I and Sudol M 1995 *Proc. Natl. Acad. Sci. U. S. A.* **92** 7819
[64] Sudol M, Bork P, Einbond A, Kastury K, Druck T, Negrini M, Huebner K and Lehman D 1995 *J. Biol. Chem.* **270** 14733
[65] Lu P J, Wulf G, Zhou X Z, Davies P and Lu K P 1999 *Nature* **399** 784
[66] Garrus J E, von Schwedler U K, Pornillos O W, Morham S G, Zavitz K H, Wang H E, Wettstein D A, Stray K M, Cote M, Rich R L, Myszka D G and Sundquist W I 2001 *Cell* **107** 55
[67] Zheng H W, You H, Zhou X Z, Murray S A, Uchida T, Wulf G, Gu L, Tang X R, Lu K P and Xiao Z X J 2002 *Nature* **419** 849
[68] Gao J, Wang L, Kang S G, Zhao L, Ji M, Chen C, Zhao Y, Zhou R and Li J 2014 *Nanoscale* **6** 12828

[69] Chin D and Means A R 2000 *Trends Cell Biol.* **10** 322
[70] Kuboniwa H, Tjandra N, Grzesiek S, Ren H, Klee C B and Bax A 1995 *Nat. Struct. Biol.* **2** 768
[71] Shen X C, Valencia C A, Szostak J, Dong B and Liu R H 2005 *Proc. Natl. Acad. Sci. U. S. A.* **102** 5969
[72] Ikura M, Clore G M, Gronenborn A M, Zhu G, Klee C B and Bax A 1992 *Science* **256** 632
[73] Crivici A and Ikura M 1995 *Annu. Rev. Biophys. Biomol. Struct.* **24** 85
[74] Brokx R D, Lopez M M, Vogel H J and Makhatadze G I 2001 *J. Biol. Chem.* **276** 14083
[75] Yang C, Jas G S and Kuczera K 2004 *Biochim. Biophys. Acta* **1697** 289
[76] Novoselov K S, Geim A K, Morozov S V, Jiang D, Katsnelson M I, Grigorieva I V, Dubonos S V and Firsov A A 2005 *Nature* **438** 197
[77] Feng L Z and Liu Z A 2011 *Nanomedicine* **6** 317
[78] Zuo G, Zhou X, Huang Q, Fang H and Zhou R 2011 *J. Phys. Chem. C* **115** 23323
[79] Tu Y, Lv M, Xiu P, Huynh T, Zhang M, Castelli M, Liu Z, Huang Q, Fan C, Fang H and Zhou R 2013 *Nat Nanotechnol* **8** 594
[80] Li Y, Yuan H, von dem Bussche A, Creighton M, Hurt R H, Kane A B and Gao H 2013 *Proc. Natl. Acad. Sci. U. S. A.* **110** 12295
[81] Chen J Y, Wang D L, Xi J F, Au L, Siekkinen A, Warsen A, Li Z Y, Zhang H, Xia Y N and Li X D 2007 *Nano Lett.* **7** 1318
[82] Chithrani B D, Ghazani A A and Chan W C W 2006 *Nano Lett.* **6** 662
[83] Qiu Y, Liu Y, Wang L M, Xu L G, Bai R, Ji Y L, Wu X C, Zhao Y L, Li Y F and Chen C Y 2010 *Biomaterials* **31** 7606
[84] Maiorano G, Sabella S, Sorce B, Brunetti V, Malvindi M A, Cingolani R and Pompa P P 2010 *ACS Nano* **4** 7481
[85] Hu W B, Peng C, Lv M, Li X M, Zhang Y J, Chen N, Fan C H and Huang Q 2011 *ACS Nano* **5** 3693
[86] Lesniak A, Fenaroli F, Monopoli M R, Aberg C, Dawson K A and Salvati A 2012 *ACS Nano* **6** 5845
[87] Wang F J, Yu L, Monopoli M P, Sandin P, Mahon E, Salvati A and Dawson K A 2013 *Nanomedicine* **9** 1159
[88] Wang L M, Liu Y, Li W, Jiang X M, Ji Y L, Wu X C, Xu L G, Qiu Y, Zhao K, Wei T T, Li Y F, Zhao Y L and Chen C Y 2011 *Nano Lett.* **11** 772
[89] Larson T A, Joshi P R and Sokolov K 2012 *ACS Nano* **6** 9182
[90] Walkey C D, Olsen J B, Guo H B, Emili A and Chan W C W 2012 *J. Am. Chem. Soc.* **134** 2139
[91] Deng Z J, Liang M T, Monteiro M, Toth I and Minchin R F 2011 *Nat. Nanotechnol.* **6** 39
[92] Wang L, Li J, Pan J, Jiang X, Ji Y, Li Y, Qu Y, Zhao Y, Wu X and Chen C 2013 *J. Am. Chem. Soc.* **135** 17359
[93] Rompel A, Cinco R M, Latimer M J, McDermott A E, Guiles R D, Quintanilha A, Krauss R M, Sauer K, Yachandra V K and Klein M P 1998 *Proc. Natl. Acad. Sci. U. S. A.* **95** 6122
[94] Zhang F C 2003 *Phys. Rev. Lett.* **90** 207002
[95] Chai Z F, Zhang Z Y, Feng W Y, Chen C Y, Xu D D and Hou X L 2004 *J. Anal. At. Spectrom.* **19** 26
[96] Qu Y, Li W, Zhou Y L, Liu X F, Zhang L L, Wang L M, Li Y F, Iida A, Tang Z Y, Zhao Y L, Chai Z F and Chen C Y 2011 *Nano Lett.* **11** 3174
[97] Chen C Y, Li Y F, Qu Y, Chai Z F and Zhao Y L 2013 *Chem. Soc. Rev.* **42** 8266

[98] Vairavamurthy A 1998 *Spectrochim. Acta, Pt. A: Mol. Biomol. Spectrosc.* **54** 2009

[99] Mcrae R, Lai B, Vogt S and Fahrni C J 2006 *J. Struct. Biol.* **155** 22

[100] Bohic S, Murphy K, Paulus W, Cloetens P, Salome M, Susini J and Double K 2008 *Anal. Chem.* **80** 9557

[101] Corezzi S, Urbanelli L, Cloetens P, Emiliani C, Helfen L, Bohic S, Elisei F and Fioretto D 2009 *Anal. Biochem.* **388** 33

[102] Zheng J, Zhang C W and Dickson R M 2004 *Phys. Rev. Lett.* **93** 077402

[103] Zheng J, Nicovich P R and Dickson R M 2007 *Annu. Rev. Phys. Chem.* **58** 409

[104] Guo W W, Yuan J P and Wang E K 2009 *Chem. Commun.* 3395

[105] Huang C C, Yang Z, Lee K H and Chang H T 2007 *Angew. Chem. Int. Ed.* **46** 6824

[106] Luedtke W D and Landman U 1996 *J. Phys. Chem.* **100** 13323

[107] Le Guevel X, Hotzer B, Jung G, Hollemeyer K, Trouillet V and Schneider M 2011 *J. Phys. Chem. C* **115** 10955

[108] Padovan-Merhar O, Lara F V and Starr F W 2011 *J. Chem. Phys.* **134** 244701

[109] Sun C J, Yuan Y, Xu Z H, Ji T J, Tian Y H, Wu S, Lei J L, Li J Y, Gao N and Nie G J 2015 *Bioconjugate Chem.* **26** 193

[110] Sun C J, Yang H, Yuan Y, Tian X, Wang L M, Guo Y, Xu L, Lei J L, Gao N, Anderson G J, Liang X J, Chen C Y, Zhao Y L and Nie G J 2011 *J. Am. Chem. Soc.* **133** 8617

[111] Wang Y L, Cui Y Y, Liu R, Wei Y T, Jiang X G, Zhu H R, Gao L, Zhao Y L, Chai Z F and Gao X Y 2013 *Chem. Commun.* **49** 10724

[112] Liu R, Wang Y L, Yuan Q, An D Y, Li J Y and Gao X Y 2014 *Chem. Commun.* **50** 10687

[113] Soderberg A, Sahaf B and Rosen A 2000 *Cancer Res.* **60** 2281

[114] Fang J G, Lu J and Holmgren A 2005 *J. Biol. Chem.* **280** 25284

[115] Berggren M, Gallegos A, Gasdaska J R, Gasdaska P Y, Warneke J and Powis G 1996 *Anticancer Res.* **16** 3459

[116] Kahlos K, Soini Y, Saily M, Koistinen P, Kakko S, Paakko P, Holmgren A and Kinnula V L 2001 *Int. J. Cancer* **95** 198

[117] Chen R, Li L and Weng Z P 2003 *Proteins: Struct. Funct. Bioinform.* **52** 80

[118] An D, Su J, Weber J K, Gao X, Zhou R and Li J 2015 *J. Am. Chem. Soc.* at press (doi:10.1021/jacs.5b00888)